\documentclass[twocolumn,superscriptaddress,amssymb,aps]{revtex4}

\usepackage{graphicx,dcolumn,bm,amsmath,color,bm}

\newcommand{\eq}[1]{Eq.~(\ref{#1})}
\newcommand{\fig}[1]{Fig.~\ref{#1}}

\newcommand{\eeq}{ \end{equation} }
\newcommand{\beq}{ \begin{equation} }

\newcommand{\bra}{ {\bf r}_1 }
\newcommand{\brb}{ {\bf r}_2 }
\newcommand{\brr}{ {\bf R} }
\newcommand{\bdr}{ \Delta {\bf  r} }
\newcommand{\bfr}{ {\bf r} }

\newcommand{\bz}{ {\bf \hat{z}} }
\newcommand{\bn}{ {\bf \hat{n}} }
\newcommand{\bx}{ {\bf \hat{x}} }

\newcommand{\bw}{   \hat{\omega}}
\newcommand{\bwa}{   \hat{\omega}_{1} }
\newcommand{\bwb}{   \hat{\omega}_{2} }

\newcommand{\nab}{  {\bf \nabla }}

\begin{document}

\title{Twisting with a twist: supramolecular helix fluctuations in chiral nematics}

\author{H. H. Wensink and C. Ferreiro-C\'{o}rdova}
\email{wensink@lps.u-psud.fr}
\affiliation{Laboratoire de Physique des Solides, CNRS, Universit\'e Paris-Sud, Universit\'{e} Paris-Saclay,  91405 Orsay, France}

\begin{abstract} Most theoretical descriptions of lyotropic cholesteric liquid crystals to date  focus on homogeneous systems in which the rod concentration, as opposed to the rod orientation, is uniform. In this work, we build upon the Onsager-Straley theory for twisted nematics and study the effect of weak concentration gradients, generated by some external potential,  on the cholesteric twist. We apply our theory to chiral nematics of nanohelices in which the supramolecular helix sense is known to spontaneously change sign upon variation of particle concentration, passing through a so-called compensation point at which the mesoscopic twist vanishes. We show that the imposed field offers exquisite control of the handedness and magnitude of the helicoidal director field, even at weak field strengths.   Within the same framework we also quantify the director fluctuation spectrum and find evidence for a correlation length diverging at the compensation point.
\end{abstract}

\maketitle

\section{Introduction}

Chiral intermolecular forces are essential for stabilizing the building blocks of life (e.g. the amino acids that make up DNA) and play an important role in key biological processes.  Condensed phases composed of  chiral constituents exhibit a much richer phase morphology than their non-chiral counterparts. Examples are liquid crystal mesophases consisting of elongated chiral mesogens which may form twisted nematic \cite{borsch2013} or cubic blue phases \cite{coles2005} whose chirality-induced periodic mesostructure endows them with special opto-electronic properties \cite{gennes-prost}. These materials find important applications in electronic displays, smart windows, optical switches, photonics and cosmetic products. 

Chiral nanoparticles are ubiquitous in the biological realm. Examples of chiral biopolymers capable of forming (chiral) liquid crystals include DNA \cite{livolant1996}, chitin \cite{belamie2004}, collagen \cite{giraud1992,giraud2008a},  cellulose \cite{gray-cullulose,heux2000}, phytosterol \cite{rossi2015}, and filamentous {\em fd} virus particles \cite{dogic2000a,grelet-fraden_chol}.  Some of these systems currently witness an active field of experimental research in which the role of biomolecular chirality on the mesoscopic material properties in relation to possible applications as functional materials is being extensively explored \cite{shopsowitz2010,chung2011,hamley2010, lagerwall2014a,usov2015}. 

\begin{figure}[h]
\centering
\includegraphics[width= 0.9 \columnwidth]{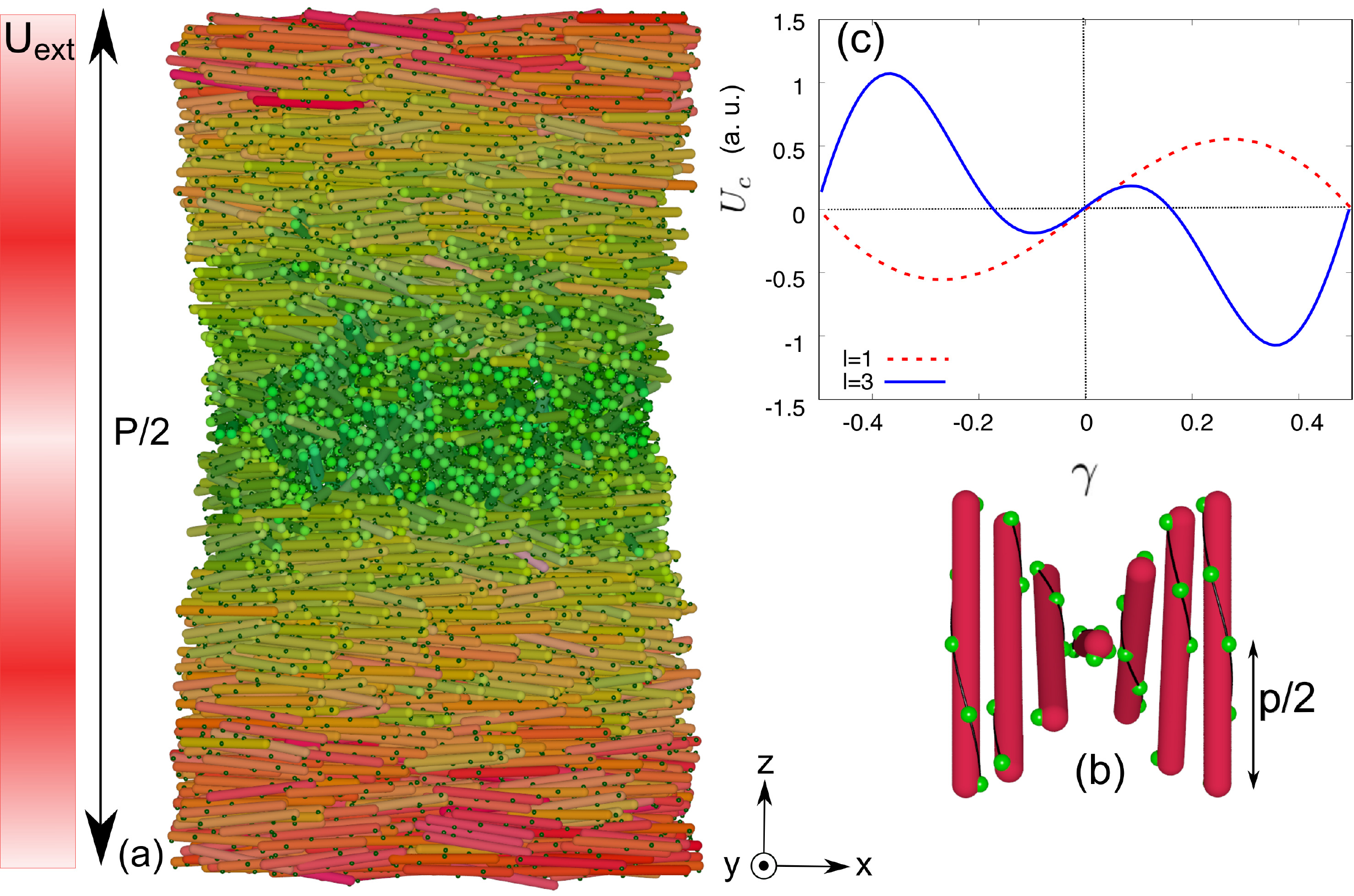}
\caption{ \label{fig0} (a) Simulation snapshot of a left-handed (LH) cholesteric phase of helical patchy cylinders. The vertical system dimension corresponds to half the pitch  $P$ of the helical director field (rod orientations are color coded).     Reprinted from Ref. \cite{ruzicka_sm2016}.  An external field $U_{\rm ext}$ coupling to the local particle concentration creates a non-uniform density along the pitch direction $z$. (b)  The particles consist of a soft helical potential  (indicated by the green dots) with pitch $p$ wrapped around the surface of a cylindrical hard core (in red). (c) The mean-field chiral potential between a rod pair at fixed centre-of-mass distance depends on the interrod angle $\gamma$ and may display a single-mimimum ($\ell = 1$) or double-minimum behavior ($\ell = 3$), depending on the sign and amplitude of the molecular pitch $p$.  Helices with $\ell =1$  are LH, those with $\ell =3$ possess a RH symmetry \cite{wensink_epl2014}.  } 
\end{figure}

Recent theoretical and simulation studies utilizing coarse-grained models for curled hard cylinders  \cite{kolli2014b,dussi2015a,wensink2015a} or helical patchy rods \cite{emelyanenko2000,wensink_epl2014,ruzicka_sm2016,kuhnhold_jcp2016} have shed new light on how molecular chirality translates into various macroscopic structures. Most of the focus has been on cholesteric liquid crystals. These structures are essentially nematic (no long-range positional order) but the local director exhibits a helical precession, characterized by an intrinsic length scale, the helical pitch $P$, and handedness (left-handed, LH or right-handed RH, see \fig{fig0}). 
One of the remarkable findings emerging from these studies is that the cholesteric sense (a left- or right-handed twist) is not only dictated by the chirality at the particle scale \cite{katsonis2012a} but also by the thermodynamic state of the system  \cite{dussi2015a}.  Helical mesogens with a certain prescribed molecular helicity may undergo spontaneous sense inversions by subtle variations of the overall particle concentration, pressure or temperature \cite{wensink_epl2014}.  Temperature-induced sense inversions are not uncommon in certain thermotropic systems \cite{slaney1992,yamagishi1990,toriumi1983,watanabe1988,stegemeyer1989},
but their origin is unclear. Most likely,  subtle modifications in the molecular chirality or solvent conditions upon variation of  temperature are at the core of these trends.  The supramolecular handedness may also be controlled using photosensitive chiral dopants \cite{mathews2010}.  Furthermore, mixing components each with a different sign and magnitude of the molecular chirality may  lead to situations where the global twist vanishes. These particular states are usually referred to as compensated or racemic cholesterics \cite{muller1973} and re-emerge in nanohelix cholesterics at the inversion point where the handedness changes sign and the supramolecular twist becomes  zero \cite{lubensky1996}.  The lyotropic case is surprising in that spontaneous sense inversions happen at {\em fixed} internal chirality and interaction range. The inversions are brought about solely by a subtle interplay between concentration and (local) particle alignment \cite{dussi2015a,wensink_epl2014}.  A further experimental exploration of these sense inversions, which have to date not been identified in the biofibril suspensions listed above,  is desirable as it may open up new possibilities to tune the optical properties of lyotropic cholesteric materials \cite{mitov2016}.

In this work we take a closer look at fluctuations in the supramolecular twist in case the particle concentration is no longer spatially uniform but subject to a weak modulation induced by some external field. The main questions we set out to address are the following. First: How does a weak modulation of the particle concentration couple to the local director twist and can we exploit this to generate more complex non-uniform periodic twist profiles?  The second question relates to fluctuations in the supramolecular twist induced by thermal motion;  Is there a diverging length scale associated with director  fluctuations upon approach of the compensation point, and, if so, how does this correlation length depend on particle concentration?

To address both issues we revisit Onsager's second-virial theory \cite{Onsager49} for nematic phases of slender hard rods, supplemented with Straley's extension \cite{straley76} to account for the effect of a non-uniform (e.g. twisted) director field. We further generalize the framework toward systems with a weak gradient in the particle concentration. The ensuing theory is essentially a hybrid square-gradient theory accounting for the subtle coupling between concentration and director deformations mediated by the local particle orientations.  An important advantage of using the Onsager-Straley approach, in contrast to some of the more expansive density functional theories formulated for chiral nematics \cite{belli2014a, allen2016a,tortora2017},  is the direct connection with the pair potential of the (helical) nanorods. No experimental or simulation input is required  to quantify the elastic properties of the system since they are intrinsically calculable within the theory. The approach thus enables us to predict the fluctuation spectrum of nanohelix cholesterics on a microscopic footing. 


The  manuscript is organized as follows. We begin in Section II by laying out a simple square-gradient formalism derived from the Onsager-Straley theory for chiral nematics. The required microscopic input parameters are discussed in Section III based on a rigid hard rod model supplemented with some tractable helical potential mimicking the twist propensity of a pair of soft helical filaments. The implications of a weak concentration modulation along the pitch axis on the local director twist is investigated in Section IV and the director fluctuation spectra will be analyzed and discussed in detail in Section V. The main conclusions drawn from this study will be formulated in the final Section.

\section{Cholesteric systems with non-uniform particle concentration}
The starting point of our analysis is Onsager's classical second-virial theory \cite{Onsager49}  designed for fluid phases of infinitely slender, rigid filaments where interactions involving more than two particles are highly improbable.   The excess free energy  $F_{ex}$ in units of the thermal energy $k_{B}T$ (with temperature $T$ and Boltzmann's constant $k_{B}$) may be generalized for inhomogeneous systems and formally reads  \cite{poni1988, allenevans}
 \beq
\frac{ F_{ex}[\rho]}{k_{B}T}  = -\frac{1}{2}  \iint d \bra d \brb \langle \langle   \rho(\bra, \bwa \cdot \bn(\bra)  ) \rho(\brb, \bwb \cdot \bn(\brb) ) \Phi \rangle \rangle,
 \eeq
where the Mayer function $\Phi = e^{- U/k_{B}T} -1$ relates to  the pair potential $U$ between two particles (``1'' and ``2''). It depends explicitly on their mutual orientation, indicated by the unit vectors $\bw_{i}$,  and their centre-of-mass distance $\bra - \brb$. The one-body density $\rho$ expresses the probability to find a rod with centre-of-mass at position $\bfr$ and orientation $\bw$ with respect to a spatially varying director field $\bn(\bfr)$.  Brackets denote a double angular average $\langle \langle  \cdot \rangle \rangle = \int d \bwa \int d \bwb $. Our working assumption is that gradients in the nematic director as well as in the particle concentration extend over distances far greater than the typical particle scale. Defining new coordinates $\brr = (\bra + \brb)/2$ and $\bdr = \bra - \brb$ we may expand  $\rho$ up to linear order in $\Delta \bfr $. This  yields two gradient contributions, one for the concentation and  a second one describing spatial variations of the director field \cite{straley76}, respectively
\begin{align}
 \rho(\bfr_{i}, \bn(\bfr_{i}) \cdot \bw_{i} )  &= \rho(\brr,  \bw_i \cdot \bn(\brr)  )  \nonumber \\ 
& \pm \left ( \frac{\bdr}{2} \cdot \nabla_{\brr} \right )  \rho (\brr , \bw_i \cdot  \bn(\brr)  ) \nonumber \\ 
& \pm \left ( \frac{\bdr}{2} \cdot \nabla_{\brr} \right ) \left ( \bn(\brr) \cdot \bw_{i} \dot{\rho} (\brr,  \bw_{i} \cdot \bn(\brr) ) \right ),  
\end{align}
for $i=1(+),2(-)$, in terms of the partial derivative of the one-body density with respect to orientation $  \dot{\rho} (\brr,  \bw \cdot \bn(\brr) ) = \partial \rho (\brr,  \bw \cdot \bn(\brr) ) / \partial (\bw \cdot  \bn(\brr)  ) $.  In the following we shall focus on a weakly twisted director field  with a helical axis fixed along the $z-$direction of the laboratory frame which we denote by Cartesian coordinates $(X,Y,Z)$. The twist deformation then reads $\bn(Z) \approx (1 , \varphi(Z), 0)$ with a non-uniform twist angle $\varphi(Z)$ ($| \nabla \varphi| \ll 1$).  An expansion of the free energy per unit surface $A$ up to second order in the gradients gives
\begin{widetext}
\begin{align}
& \frac{F_{ex}}{A k_{B}T} =  \int d Z \left   \langle  \left  \langle \left \{  \frac{1}{2}M_{0} \rho( Z,\bwa ) \rho(Z, \bwb  )  +  \frac{1}{2} M_{1} \rho(Z, \bwa ) \omega_{2y} \dot{\rho} (Z, \bwb ) \nabla \varphi(Z)    \right . \right . \right . \nonumber \\
& \left . \left . \left .  + \frac{1}{4}  M_{2} \left [ \nabla   \rho( Z, \bwa  )  \nabla   \rho(  Z, \bwb  )  +   \omega_{1y} \omega_{2y} \dot{\rho} ( Z, \bwa ) \dot{\rho} (Z, \bwb ) ( \nabla   \varphi(Z))^{2}  \right ]  \right \} \right \rangle \right \rangle.
\label{fsg_simple}
\end{align}
\end{widetext}
\noindent In deriving the above, we have imposed mirror reflection symmetry, $\rho(Z, \bw) = \rho(-Z, \bw)$ by requiring that all linear terms $\nabla \rho$ be zero.
The kernels $M_{n}$ refer to the $n$-th moment of the Mayer function and are defined as
\beq
M_{n} (\bwa, \bwb) = -\int d \bdr (\bdr \cdot  \bz)^{n} \Phi(  \bdr   , \bwa , \bwb ).
\label{kernelm}
\eeq
These quantities depend explicitly on the mutual particle orientation of a rod pair and provide the key microscopic input of our theory. The kernels will be specified in the next Section.
The odd term $M_{1}$ is only non-zero if the rod interactions are {\em chiral} in which case the direction of twist deformation matters, i.e.,  $\nabla \varphi \neq - \nabla \varphi$. For achiral particles all terms linear in $\nabla \varphi$  vanish.  The excess term involves pair-interactions only and is merely approximate at elevated particle densities.  The remaining free energy contributions on the other hand are  exact and represent the free energy of an ideal gas of rodlike particles via
\beq
 \frac{F_{id}[\rho] }{A k_{B}T} =  \int d Z   \langle  \rho(Z,  \bw ) [ \ln   {\mathcal V} \rho(Z,  \bw ) - 1 +   U_{ext} (Z, \bw) ] \rangle,
 \label{fideal}
\eeq
where the last term imparts the effect of some externally imposed potential $U_{ext}$ and ${\mathcal V}$ is an immaterial thermal volume containing contributions from the rotational momenta of the particles.
The next step is to minimize the {\em total} free energy with respect to the density $\rho(Z, \bw)$ while assuming the density to be unaffected by the weak director twist. This is done by means of a functional minimization $ \frac{\delta } {  \delta \rho(Z, \bw)}\left  [F - \mu \int d Z \langle \rho(Z, \bw) \rangle \right ] _{\nabla \varphi =0 } =0 $, in terms of  a chemical potential $\mu$ acting as a Lagrange multiplier to ensure a fixed particle number; $\int d Z \langle \rho(Z, \bw) \rangle = N/A$. The result is an Euler-Lagrange equation for the one-body density which can be recast as a Boltzmann exponent
\begin{align}
 \rho(Z,  \bwa)  &=\frac{1}{{\mathcal V}}  \exp (  -  \beta [ U_{\rm S} (Z, \bwa) - \mu ]), 
 \label{minro}
 \end{align}
 in terms of a self-consistent field $U_{\rm S}$ combining some effective internal potential due to rod-rod correlations and the external one
 \beq
U_{\rm S}(Z,\bwa)  =   \langle  M_{0} \rho(Z,  \bwb  ) + \frac{1}{4} M_{2} \nabla^{2} \rho(Z,  \bwb  ) \rangle_{\bwb} + U_{\rm ext} (Z, \bwa).   
 \eeq
Minimization of the total free energy with respect to the twist deformation $\delta F/\delta \nabla \varphi(Z) =0$ yields for the equilibrium twist
\beq
 \nabla \varphi (Z) = \frac{K_{t}(Z)}{ K_{2}(Z)} 
\label{elphi}
\eeq
where the coefficients relate to a weighted double angular average of the kernels
\begin{align}
\beta K_{t}(Z) & = -\frac{1}{2}  \langle \langle M_{1} w_{2y} \rho(Z,  \bwa  )\dot{\rho}(Z,  \bwb  )  \rangle  \rangle, \nonumber \\ 
\beta K_{2}(Z) &=  \frac{1}{2} \langle \langle w_{1y} w_{2y}  M_{2} \dot{\rho}(Z,  \bwa  )\dot{\rho}(Z,  \bwb  ) \rangle \rangle. 
\label{coefa}
\end{align}
The results for systems with a {\em uniform} particle concentration $\rho_{0}$ are easily retrieved by setting $\rho(Z, \bw) = \rho_{0} f_{0}(\bw)$. The local orientation distribution function (ODF) $f_{0}$ then follows from $f( \bwa)  = {\mathcal N} \exp \left ( -  \rho_{0} \langle  M_{0}(\bwa, \bwb) f_{0} ( \bwb  ) \rangle_{\bwb}  \right )$ with the constant ${\mathcal N}$ ensuring normalization via $\langle  f_{0}(\bw) \rangle =1 $.  Likewise, the two coefficients \eq{coefa} reduce to the familiar torque-field and the (Frank) twist elastic constants, defined as \cite{allenevans}
\begin{align}
\beta K_{t} & = -\frac{\rho_{0}^{2}}{2}  \langle \langle M_{1} w_{2y} f_{0} (\bwa ) \dot{f}_{0} ( \bwb  )  \rangle  \rangle, \nonumber \\ 
\beta K_{2} &=  \frac{\rho_{0}^{2}}{2} \langle \langle w_{1y} w_{2y}  M_{2} \dot{f}_{0}(\bwa) \dot{f}_{0} ( \bwb  ) \rangle \rangle. 
\label{k2}
\end{align}
The ratio of these two give a uniform twist deformation  $\nabla \varphi(Z) = K_{t}/K_{2} = q_{0}$  with $q_{0}$  a wavenumber inversely proportional to the pitch of the cholesteric system.  The more general expressions \eq{minro} and \eq{coefa}  enable us to compute the non-uniform twist profile of a cholesteric phase exposed to an  external potential acting along the pitch direction. In  Section IV, we shall take a closer look at the implications of a weak concentration gradients imposed by some arbirtrary external field  (related to e.g. particle sedimentation, solvent evaporation, or the presence of a substrate). But first, we need to specify the microscopic interactions that underpin the stability of cholesteric order in suspensions of helical filaments.

\section{Coarse-grained potential for rigid helical filaments}

\begin{figure}
\centering
\includegraphics[width= 0.75 \columnwidth]{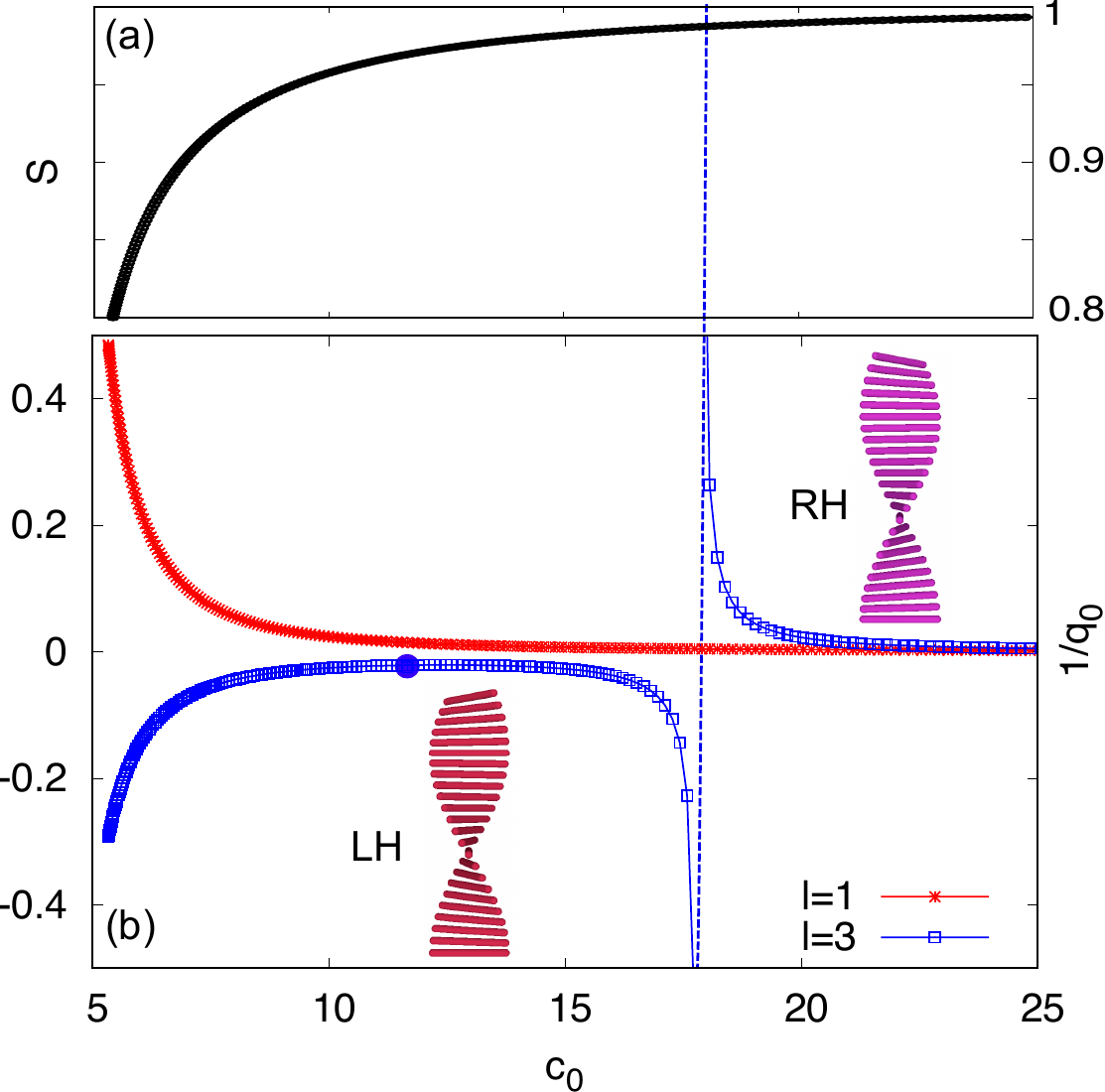}
\caption{ \label{fig1} (a) Local nematic order parameter $S$ versus concentration for a lyotropic cholesteric of chiral rods. (b) Corresponding helical pitch (in units $L/\bar{\varepsilon}_{c}$) and handedness for a system with a monotonically decreasing pitch ($\ell =1$) and a system exhibiting a spontaneous inversion of the cholesteric handedness ($\ell =3$). At the compensation point ($c_{0} \approx 18$, $S \approx 0.98$) the supramolecular twist vanishes as indicated by a divergence of the pitch (vertical dotted line).  } 
\end{figure}

Let us consider the interactions between a pair of hard cylindrical rods with length $L$ and diameter $D$, each padded with some helical surface pattern, resembling a helical `patchy' particle \cite{ruzicka_sm2016,kuhnhold_jcp2016}. For reasons of symmetry, the even kernels $M_{0}$ and $M_{2}$ featuring in the square-gradient free energy  \eq{fsg_simple} only depend on the achiral hard cores. The Mayer function $\Phi$ yields -1 when the cores overlap and zero otherwise. For hard cylinders with infinite length-to-width ratio $L/D \rightarrow  \infty$ the kernels  correspond to the following (generalized) excluded volumes \cite{odijkelastic, wensinkjackson}
\begin{align}
M_{0} & \sim  2L^{2}D | \sin \gamma  |,  \nonumber \\ 
M_{2}  & \sim  \frac{1}{6} L^{4}D  | \sin \gamma  |  [ (\bwa \cdot \bz)^{2} + (\bwb \cdot \bz)^{2} ], 
\end{align} 
with $| \sin \gamma |= | \bwa \times  \bwb |$. The odd kernel $M_{1}$ depends on the specific chiral interaction $U_{c}$ between the helical filaments and is strictly zero in the absence of chirality. For weakly chiral interactions ($U_{c}  \ll k_{B}T$) it is justified to approximate $\Phi \approx - \beta U_{c}$.  To mimic the effective potential between soft helical filaments \cite{wensink_epl2014} we propose the following simplified form
\beq
U_{c} \sim \varepsilon_{c} g(\Delta r) (\bwa \times \bwb  \cdot \Delta {\bf  r} ) 
\begin{cases} 
       \frac{\pi}{2 \gamma_{\rm c}}    \cos \left (  \frac{\pi}{2} \frac{ \ell \gamma}{\gamma_{\rm c}} \right )   &  | \gamma |  \leq \gamma_{\rm c}  \\
     0  & | \gamma |  > \gamma_{\rm c}. 
   \end{cases}
   \label{uchir}
\eeq
This potential is intrinsically chiral since it is not invariant with respect to the inversion operation $ \Delta {\bf  r} \rightarrow -\Delta {\bf r} $. The decay  with increasing centre-of-mass distance is given by $g( \Delta r )$. The pseudoscalar form $(\bwa \times \bwb  \cdot \Delta {\bf  r} ) $ originally emerged from electric multipole expansions \cite{goossens71,meervertogenJCP}
 but has since then been consistently used in simulation models to capture chiral interactions (whether caused by quantum-mechanical or steric factors) between non-spherical mesogens   \cite{vargachiral,memmer1993,berardi_zannoni1998,germano2002a,wilson2005}.  As for the remaining parameters, $\varepsilon_{c}$ is an amplitude parameter and  $\gamma_{\rm c} $ a cut-off value for the angle, such that $U_{c}(\gamma_{\rm c})=0$. Most importantly,  $\ell = 1,3,5 \dots $ is an odd integer determining the number of local minima in $U_{c}(\gamma)$. This is illustrated in \fig{fig0}(c). The case $\ell =1$ produces a single minimum function  imparting a uniform helix sense, whereas the double-minimum form for $\ell =3$ gives rise to {\em pitch inversion} scenario where the cholesteric helix sense switches handedness upon changing the overall particle concentration of the cholesteric system. The kernel $M_{1}$ can be approximated by introducing a  cylindrical  laboratory frame $ ( \Delta r_{\perp}, \Delta z) $ 
\begin{align}
M_{1} & =  - \int d \Delta {\bf  r}  ( \Delta {\bf  r} \cdot \bz )  \beta U_{c} (  \Delta {\bf  r }   , \bwa , \bwb ) \nonumber \\
&\sim   -  \bar{\varepsilon}_{c}  L^{4} (\bwa \times \bwb  \cdot \Delta {\bf  \bz} ) \begin{cases} 
    \frac{\pi}{2 \gamma_{\rm c}}    \cos \left (  \frac{\pi}{2} \frac{ \ell \gamma}{\gamma_{\rm c}} \right )    &  | \gamma |  \leq \gamma_{\rm c}  \\
     0  & | \gamma |  > \gamma_{\rm c}, 
   \end{cases}
\label{m1chiral}
\end{align}
where the spatial integral over the decay function is now subsumed into some effective dimensionless chiral amplitude via
\beq
\bar{\varepsilon}_{c}  =  \pi \frac{\varepsilon_{c}}{k_{B}T} L^{-4} \int_{0}^{\infty}  d \Delta r_{\perp}^{2} \int_{-\infty}^{\infty} d \Delta z (\Delta z)^{2} g( \Delta r_{\perp}, \Delta z ).
\label{beps}
\eeq 
The precise form of $g(\Delta r)$ is not crucially important as long as convergence of the spatial integral is guaranteed and the condition $\bar{\varepsilon}_{c} \ll 1$ is met. 
We emphasize that the definition of $\bar{\varepsilon}_{c}$ makes the theory applicable to a wide range of cholesteric materials of rigid helical filaments where chiral forces are transmitted primarily by long-ranged, soft interactions  rather than by steric forces related to particle shape \cite{kolli2016}.
For the case $\ell=3$ the critical concentration at which a helical sense inversion occurs is inversely proportional to  $ \gamma_{\rm c}$. In our calculations, we choose  $\gamma_{\rm c} = 0.5$ in which case a pitch sense inversion occurs at a concentration of $c_{0}= \rho_{0}L^{2}D = 17.84$. The isotropic-cholesteric phase coexistence densities are located at $c_{0}^{(I)} = 4.189$ and $c_{0}^{(N)} = 5.336$ \cite{vroege92}. Some relevant numerical results for the pitch versus concentration have been compiled in \fig{fig1}. For the homogeneous systems, standard iteration routines utilizing an equidistant grid of relevant angles to discretize orientational space $\bw$ were employed to solve equations such as \eq{minro} \cite{herzfeldgrid,roij_njp2005}. 
In \fig{fig1}   two distinct scenarios are highlighted: a conventional one ($\ell =1$) in which the pitch decreases monotonically with concentration, as routinely encountered in a wide range of bio-inspired cholesteric liquid crystals  \cite{dogicfraden2000,dupreduke75,schutz2015,belamie2004,giraud2008a}. The second case ($\ell =3$)  relates to  a pitch-inversion scenario where the twist suddenly changes handedness at a critical concentration and, associated with this, a critical degree of local nematic alignment \cite{wensink_epl2014,dussi2015a,ruzicka_sm2016,kuhnhold_jcp2016}.  The microscopic underpinning for this phenomenon resides in the double-minimum form of the chiral potential (see \fig{fig0}(c)). Since the two minima are located at opposite signs of the twist angle the global twist sense imparted by the chiral potential depends critically on the  degree of nematic alignment $ \sim \langle \langle \gamma \rangle \rangle$ along the director field, which is steered by  particle concentration \cite{wensink_epl2014}.

A rough estimate for $\bar{\varepsilon}_{c}$ can be produced by assuming helical rods with some chiral charge pattern \cite{rossi2015,tombolato2006} with an effective total charge $Q_{\rm eff}$ residing on the particle surface, so that chiral forces are mediated via some screened Coulomb potential $Q_{\rm eff}^{2}\lambda_{B} \exp(-\kappa r)/r  $ with $\lambda_{B}$  the Bjerrum length and $\kappa$ the Debye screening constant related chiefly to the ionic strength of the solvent. Using this in \eq{beps} 
we write  $\bar{\varepsilon}_{c}$ as a simple product of $Q_{\rm eff}$  and a number of (dimensionless) size ratios
\beq
\bar{\varepsilon}_{c} \sim Q_{\rm eff}^{2} (\lambda_{B}/D) (D/L)^{3}(\kappa D)^{-2}. 
\eeq
We may test the usefulness of this prediction by plugging in typical numbers for e.g. filamentous virus rods \cite{tombolato2006}. Taking order-of-magnitude estimates for the relevant size ratios, $\lambda_{B}/D  \sim {\mathcal O}(10^{-1}) $, virus aspect ratio $D/L \sim {\mathcal O}(10^{-2})$, effective surface charge  $ Q_{\rm eff} \sim {\mathcal O}(10^{3})$, and electrostatic screening $\kappa D \sim {\mathcal O}(1)$, yields $\bar{\varepsilon}_{c} \sim {\mathcal O}(10^{-1})$. Similarly, reasonable estimates for cellulose nanocrystals (CNCs) \cite{schutz2015} are:  $\lambda_{B}/D  \sim {\mathcal O}(10^{-1}) $,  $D/L \sim {\mathcal O}(10^{-2}) $, $ Q_{\rm eff} \sim {\mathcal O}(10^{2})$, and $\kappa D \sim {\mathcal O}(1)$ gives $\bar{\varepsilon}_{c} \sim {\mathcal O}(10^{-1}  -10^{-2})$.  Reading off typical values in \fig{fig1}b we obtain for  the pitch length  $P \sim (2 \pi / q_{0}) ( L / \bar{\varepsilon}_{c}) \sim {\mathcal O} (L/\bar{\varepsilon}_{c})$ so that $P/L \sim {\mathcal O}(10^{1} - 10^{2})$.  Given that nanorod contour lengths lie in the range  $L\sim 0.1 -1$ microns,  the corresponding pitches amount to tens of microns, in full accordance with what is routinely measured in experiment.

\section{Impact of a weak concentration gradient along pitch direction}

Let us assume a small perturbation from the uniform particle concentration
\beq
\rho(Z, \bw) = \rho_{0} f_{0}(\bw) +  \delta \hat{\rho}_{q}(\bw) e^{iqZ},
\label{rhopert} 
\eeq
imparted by some weak external periodic potential of the form $U_{\rm ext}(Z) = \hat{u}  e^{iqZ}$ with amplitude $\hat{u} \ll 1$ acting on the positional coordinates alone. Examples could be concentration gradients imposed by e.g. an laser-optical trap, a temperature gradient, solvent evaporation or particle sedimentation or induced by the presence of a substrate or interface.  Linearising the Euler-Lagrange equation \eq{minro} we obtain a self-consistency equation for $\delta \hat{\rho}_{q}$
\beq
-\delta \hat{\rho}_{q}(\bwa) =   \rho_{0}f_{0}(\bwa)  [\beta \hat{u}  +   \langle  ( M_{0}  + \frac{q^{2}}{4} M_{2}   ) \delta \hat{\rho}_{q} (\bwb)  \rangle_{\bwb}  ],
\label{f1}
\eeq
for every mode $q \neq 0$.  Inserting the perturbed one-body density \eq{rhopert}  into the coefficients \eq{coefa} and retaining contributions up to linear order allows us to write $K_{n}(Z) \sim  K_{n}  + \delta K_{n} e^{iqZ}$ ($n=t,2$).The linear perturbations depend implicitly on particle concentration $\rho_{0}$ and wavenumber $q$ of the imposed concentration fluctuation (through \eq{f1})  and the orientational distributions via
\begin{align}
\delta K_{t} &= -\frac{\rho_{0}}{2}   [ \langle \langle M_{1} w_{2y} f_{0}(\bwa) \delta \dot{\hat{\rho}}_{q}(\bwb)  \rangle \rangle \nonumber \\ 
& +  \langle \langle M_{1} w_{2y} \delta \hat{\rho}_{q}(\bwa)  \dot{f_{0}}(\bwb)  \rangle \rangle ],    \nonumber \\ 
\delta K_{2} &= \rho_{0} \langle \langle M_{2} w_{1y}w_{2y} \delta \dot{\hat{\rho}}_{q}(\bwa)  \dot{f_{0}}(\bwb)  \rangle \rangle. 
\end{align}
The non-uniform twist then becomes up to linear order in $\delta \hat{\rho}_{q}$
\beq
 \nabla \varphi (Z)  = q_{0} +   \chi  e^{iqZ} + {\mathcal O}( \delta \hat{\rho}^{2} ),
 \label{response}
\eeq
where $q_{0} = K_{t}/K_{2}$ is the helical wave-number of the uniform cholesteric phase. 
The susceptibility $\chi = \partial q_{0}/\partial | \delta \hat{\rho}_{q} | $ has units of inverse length and expresses the non-trivial linear response of the pitch of a cholesteric nematic upon imposing a weak concentration fluctuation along the pitch direction. It reads 
\beq
\chi =   \frac{\delta K_{t} -  q_{0}\delta K_{2} }{K_{2}}, 
\eeq
and is nonzero because the local rod orientations areaffected by the imposed density gradient.  Solving \eq{f1} numerically  we find a monotonic increase of $\chi$ with the field amplitude $\hat{u}$ and a negligible dependency on $q$ in the weak-gradient regime $q \ll 1$.
\begin{figure}
\centering
\includegraphics[width= 0.75 \columnwidth]{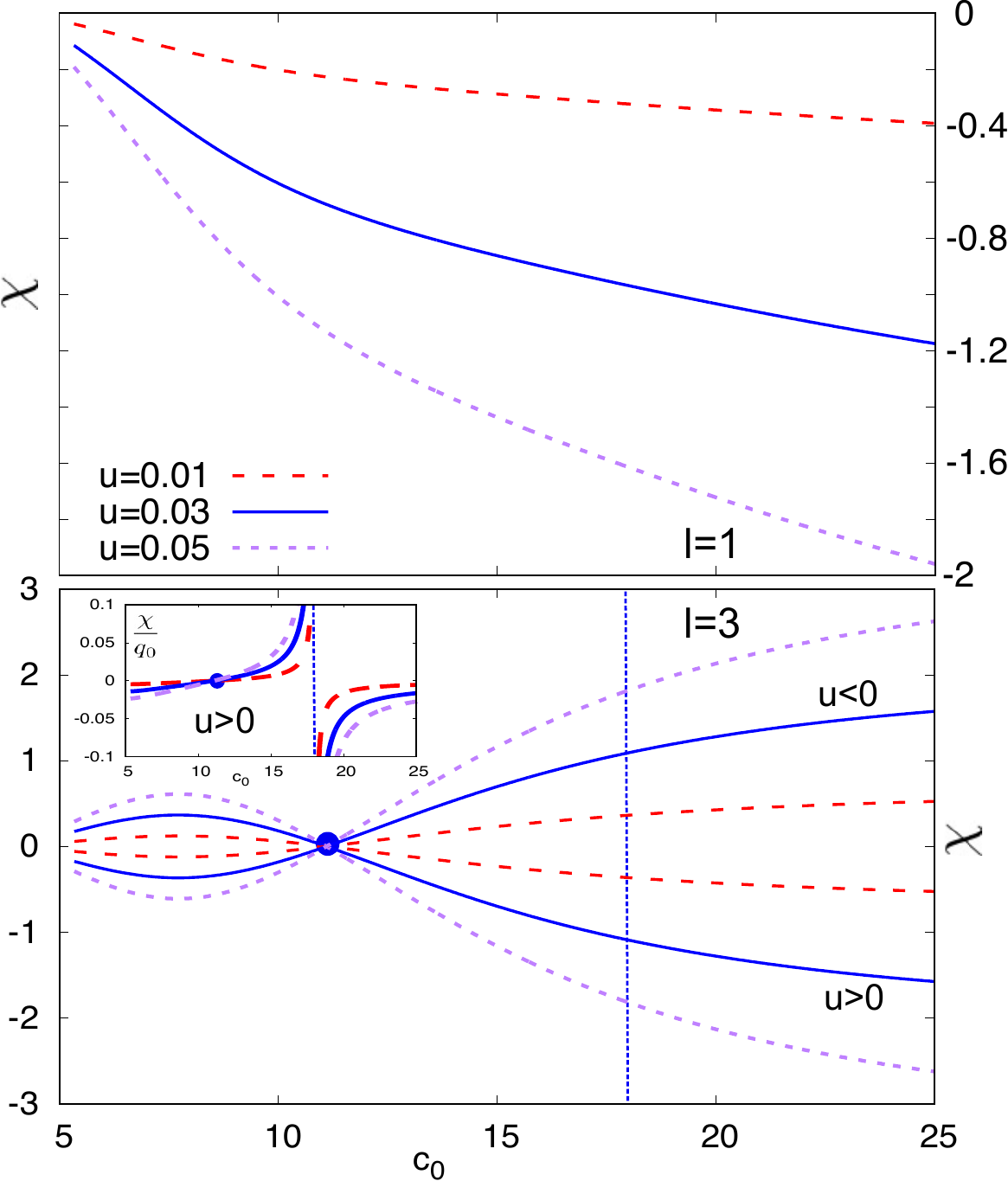}
\caption{ \label{fig2} Applying a weak external field of strength $\hat{u}$ (in units $k_{B}T$) induces a concentration modulation along the pitch axis which distorts the uniform twist of the director field. The amplitude of the local twist deformation $\chi$ (which has units inverse length, $\bar{\varepsilon}_{c}/L$) is plotted as a function of the overall particle concentration $c_{0}$.  For the case $\ell =3$ there is a point of {\em zero response} around $c_{0} \approx 11.1$ (blue dot). The compensation point where the global twist vanishes ($q_{0} \downarrow 0$) is indicated by blue vertical dotted line.  } 
\end{figure}

\eq{response} tells us that the external field renders the local twist non-uniform and causes the nematic director field to adopt a more complicated helicoidal topology.
The director component perpendicular to the reference direction ($x$-axis)  twists in the following way
 \beq
\hat{n}_{y}(Z) \approx q_{0} Z  + \chi q^{-1} \sin(qZ),  
\eeq
In practice, in view of the square-gradient approximation underpinning \eq{f1} the wavelength of the imposed concentration wave should be small ($q \ll 1$) so that  
\beq
\hat{n}_{y}(Z)  \approx  (q_{0}   + \chi)Z,  
\label{nysimple}
\eeq
independent of $q$. The evolution of the response $\chi$ as a function of the overall particle concentration is shown in \fig{fig2}.  The response is simply monotonically increasing with $c_{0}$ for the $\ell =1$ scenario (without pitch inversion), while the case $\ell = 3$ exhibits a marked point of zero response at a density preceding the compensation point.  At the zero point the effect of the applied field on the local twist vanishes. It roughly corresponds to the concentration where the derivative of the pitch with concentration becomes zero,  $\partial q_{0}/ \partial c_{0} \rightarrow 0  $ (blue dot in \fig{fig1}b). We stress, however, that the concentration-orientation coupling renders the response strongly non-linear so that $\chi$ does {\em not} obey a simple prescription $  \chi \sim \frac{\partial q_{0}}{ \partial c_{0}} \delta c_{0}(u) $, with $\delta c_{0}(u)$ the field-induced change of the local concentration, one could have naively proposed.

At the compensation point, where the intrinsic twist vanishes ($q_{0} \downarrow 0 $), a global twist can be imposed by the external field. 
Variation of the amplitude and sign of the external potential via $u$ thus allows for a judicious tuning of the handedness and the pitch length of the helicoidal director field. This is illustrated in the bottom panel of \fig{fig2}.  Typically,  an imposed field strength of $0.01 k_{B}T$  suffices to bring about a change in the helical pitch of order $\chi^{-1} \sim {\mathcal O}(L/\bar{\varepsilon}_{c})$ where $\bar{\varepsilon}_{c}$ depends on the molecular details of the filaments responsible for transmitting chirality (see \eq{beps}).  
Recalling the estimate $\bar{\varepsilon}_{c} \sim {\mathcal O}(10^{-1}  -10^{-2})$ for typical chiral nanorods (Section III) we conclude that the impact of a weak concentration gradient on the pitch is expected to be quite significant.

\section{Director fluctuations in compensated cholesterics: evidence for a diverging length-scale }

In this Section we attempt to quantify the range and strength of thermal fluctuations the helicoidal director field experiences.  We shall focus in particular on the behaviour of these fluctuations in the vicinity of the compensation point where the cholesteric twist vanishes.  In contrast to most phenomenological theories put forward to date \cite{gennes-prost,rey2010,yamashita_2004,yoshioka_epl2012}, the Onsager-Straley theory enables us to gauge the elastic properties of the cholesteric from a microscopic standpoint and establish an explicit dependence of the fluctuation spectrum with respect to particle concentration.  Let us consider the following perturbations of the helical director field
\begin{align}
\hat{n}_{x}(\brr) &=  \cos(q_{0}Z + \sum_{k_{\perp}}\delta\hat{q}_{k_{\perp}} e^{i{\bf k_{\perp}} \cdot \brr}) \nonumber \\ 
\hat{n}_{y}(\brr) &=\sin(q_{0}Z + \sum_{k_{\perp}} \delta\hat{q}_{k_{\perp}} e^{i{\bf k_{\perp}} \cdot \brr}) \nonumber \\
\hat{n}_{z}(\brr) &= \sum_{k_{\parallel}} \delta\hat{q}_{k_{\parallel}}  e^{i{\bf k_{\parallel}} \cdot \brr}, 
\label{nfluc}
\end{align}
where the amplitude $ | \delta\hat{q}_{k_{\perp}} | \ll 1$ refers to a weak modulation of the linear twist  $\varphi(Z) = q_{0}Z$ and $ | \delta\hat{q}_{k_{\parallel}} | \ll 1$ to a spatial perturbation of the pitch direction (along the $z$-axis). The change in excess free energy produced by a weak non-uniformity of the director field  takes the following form  \cite{straleychiral,wensinkjackson}
 \begin{align}
& \frac{F_{\text{twist}}}{k_{B}T}  \sim  \frac{\rho ^2}{2}  \int d \brr    \langle \langle  \int d \bdr   \partial_{\brr} (\bw_{2})  \Phi  f_{0}(\bwa)\dot{f}_{0}(\bwb) \rangle \rangle \nonumber \\
& -  \frac{\rho ^2}{4}    \int d \brr    \langle \langle  \int d \bdr   \partial_{\brr} (\bw_{1})  \partial_{\brr} (\bw_{2})  \Phi   \dot{f}_{0}(\bwa)\dot{f}_{0}(\bwb) \rangle \rangle + \cdots,
\end{align}
where $ \partial_{\brr} (\bw_{i}) =  ( \bdr  \cdot \nab_{\brr} )  \bn ({\bf R}) \cdot \bw_{i} $.  Ignoring the fluctuation terms, we easily retrieve the mean-field free energy of a weakly twisted cholesteric  by inserting \eq{nfluc} and expanding up to quadratic order in $q_{0}$  so that  $F_{\text{twist}}/k_{B}T   = -K_{t}q_{0}  + \frac{1}{2}K_{2}q_{0}^{2} $ (cf. \eq{k2}).
It is now fairly straightforward to work out the free energy change imparted by a weak spatial modulation of the helicoidal director field by inserting $\bn(\brr)$ and retaining the leading order contributions for small amplitudes $\delta \hat{q}_{k_{\perp}}$ and  $\delta \hat{q}_{k_{\parallel}}$. Focussing on the latter first,  we obtain for the free energy change associated with longitudinal director fluctuations along the pitch direction
\beq
\frac{\delta F_{\parallel}}{V} \sim \frac{1}{2} \sum_{k_{\parallel}} \left \{  \delta \hat{q}_{k_{\parallel}}^{2} \left [k_{\parallel,x}^{2} K_{3} +k_{\parallel,y}^{2} K_{2} +   k_{\parallel,z}^{2} K_{1}  \right ] \right \},
\eeq 
in terms of the splay ($K_{1}$) and bend ($K_{3}$) elastic constants, specified in the Appendix.
From the quadratic contribution we can infer the following fluctuation spectrum upon invoking the equipartition theorem \cite{gennes-prost}
 \beq
 \langle \delta \hat{q}^{2}_{k_{\parallel}} \rangle \sim \frac{k_{B}T}{V( k_{\parallel,x}^{2}K_{3} + k_{\parallel,y}^{2}K_{2} + k_{\parallel,z}^{2}K_{1})}.
 \eeq
It suggests that fluctuations in the pitch direction decay algebraically, irrespective of the cholesteric twist $q_{0}$. 
A similar analysis produces the following spectrum for the transverse fluctuations (i.e. perpendicular to the pitch axis $z$) of the local nematic director
 \beq
\langle \delta \hat{q}_{k_{\perp}}^{2} \rangle  \sim \frac{k_{B} T }{ V(k_{\perp,x}^{2} K_{3} +k_{\perp,y}^{2} K_{1} +   k_{\perp,z}^{2} K_{2}  + q_{0}^{2}  K_{*})}, 
\label{spectrans}
\eeq
where $K_{*}>0$  is an additional elastic constant specified in the Appendix. Taking the inverse Fourier transform  (FT) of this expression we find that the transverse director fluctuations along the helicoidal axis decay exponentially
\beq
\langle \delta q_{\perp}(Z)^{2} \rangle \sim \frac{k_{B}T} {V (K_{*} K_{2})^{\frac{1}{2}}}\frac{e^{- |Z|/ \xi_{z}} }{q_{0}},  
\label{decayz}
\eeq
in terms of a correlation length 
\beq
\xi_{z} \sim  \left ( \frac{K_{2}}{K_{*}q_{0}^{2}}  \right )^{\frac{1}{2}}  \sim \left ( \frac{7}{10 \pi} \right )^{\frac{1}{2}} \frac{1}{|q_{0}|c_{0}}.
\eeq
A similar behavior is found for the decay of transverse fluctuations  measured along the local director (which is fixed along the $x$-axis of the lab frame).  The approximation $k_{\perp,x}^{2} K_{3} +k_{\perp,y}^{2} K_{1} \approx k_{\perp,x}^{2} K_{3}$ seems justifiable for  concentrated hard rod systems where the splay modulus is much smaller than the bend one ($K_{1} \ll K_{3}$). Performing an inverse FT of \eq{spectrans} we obtain a similar exponential form $\langle \delta q_{\perp}(X)^{2} \rangle \sim k_{B}T V^{-1} (K_{*} K_{3})^{-\frac{1}{2}}
e^{- |X|/ \xi_{x}} /q_{0}  $ whose amplitude now involves the bend modulus $K_{3}$. The correlation length for transverse director fluctuations probed along the local director also diverges at the compensation point, albeit with a different concentration scaling than $\xi_{z}$
\beq
\xi_{x} \sim  \left ( \frac{K_{3}}{K_{*}q_{0}^{2}}  \right )^{\frac{1}{2}}  \sim  \frac{1}{5^{\frac{1}{2}}} \frac{1}{|q_{0}|}.
\eeq
\begin{figure}
\centering
\includegraphics[width= 0.75 \columnwidth]{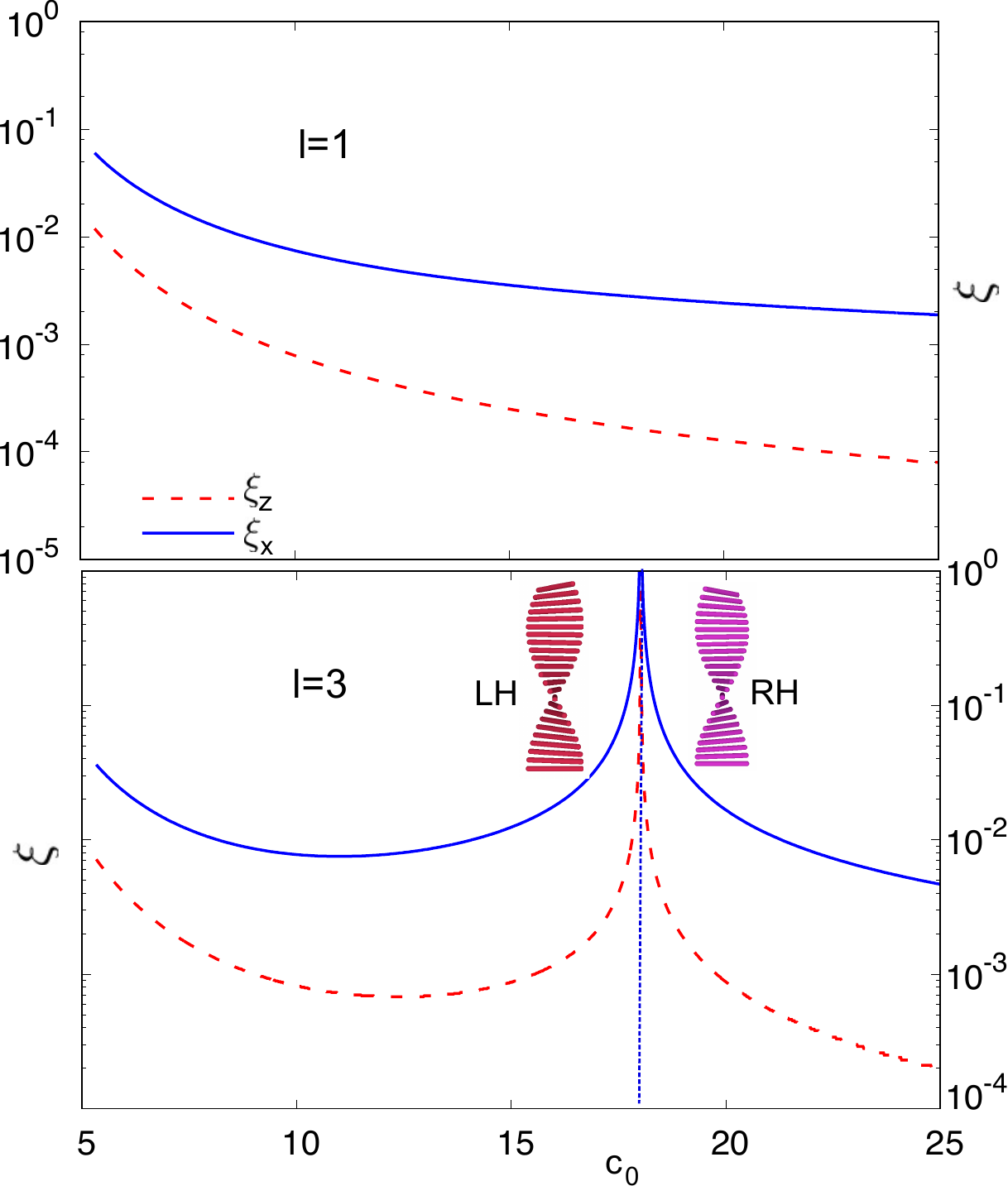}
\caption{ \label{fig3}  Correlation length (in units $L/\bar{\varepsilon}_{c}$) measuring the decay of director fluctuations transverse to the pitch axis  probed along the pitch axis ($\xi_{z}$) and along the local nematic director ($\xi_{x}$). For the case $\ell =3$ both length scales diverge at the compensation point where the global twist vanishes. } 
\end{figure}
\noindent This correlation length is of the order of the helical pitch  $1/q_{0}$ whereas  $\xi_{z} < \xi_{x}$ throughout the probed concentration range.  The concentration dependence of these correlation lengths can be established in explicit form from the asymptotic results for the elastic constants of infinitely slender hard rods which have been compiled in the Appendix. 
The expressions above clearly demonstrates that both correlation lengths and their respective amplitudes diverge at the compensation point (see \fig{fig3}).  This suggests that the crossover from one handedness to the other  upon changing the thermodynamic state (particle concentration or temperature) as reported in a number of recent studies \cite{wensink_epl2014,belli2014a,dussi2015a,ruzicka_sm2016,kuhnhold_jcp2016} constitutes some higher-order phase transition where director fluctuations diverge critically at the compensation point.

\section{Conclusions}

We have investigated in which way the supramolecular twist in a lyotropic cholesteric structure is affected by weak gradients in particle concentration as well as by thermal fluctuations. Our focus is on lyotropic assemblies of helical nanohelices where chiral torques are transmitted through some weak helical surface potential for which we propose a simple coarse-grained potential. This serves as the microscopic basis of an Onsager-Straley theory for twisted nematics which we have generalized to account for weak concentration gradients.  
Applying a generic external potential acting only on the centre-of-mass coordinates induces a weak modulation of the concentration along the pitch direction. We show that the concentration gradients couple non-linearly to the cholesteric twist via the average rod orientations and  demonstrate that spatially non-uniform twist patterns can be generated in this manner. 
In case the system is near a so-called {\em compensation point} where the global twist  but not the molecular chirality vanishes, a significant change in the pitch can be realized for weak potential amplitudes. This effect can be exploited to tune  the supramolecular twist of lyotropic materials without the need to modify the molecular chirality, for instance, by changing the solvent conditions or temperature.   

In the second part of this work we use the Onsager-Straley framework to identify how the twisted director field is affected by thermal fluctuations. Upon deriving the director fluctuation spectrum for nanohelix cholesterics we put forward an analytical expression relating the correlation length which measures the decay of the local director fluctuations along and transverse to the pitch axis to the microscopic properties of the constituents. We show that this correlation length diverges at the compensation point where the global twist vanishes.

From an experimental point of view, it would be highly desirable to dispose of model systems in which the molecular chirality (e.g. the microscopic pitch) can be carefully controlled. These would facilitate a systematic investigation of the relation between the micro- and mesoscale chirality and identify the presence of compensation points, cholesteric sense inversions and non-monotonic trends in the pitch versus particle concentration. Interesting opportunities lie in the application of filamentous phages to generate rod-shaped particles with tunable persistence length and chirality \cite{barry2009,dogic2016}, or in the self-assembly of chiral fibres of stacked organic compounds \cite{engelkamp1999} or inorganic nanoparticles with bespoke shape and interactions \cite{gao2011}.

\section*{Acknowledgements}

The authors acknowledge helpful discussions with Paul van der Schoot. This work was funded by a Young Researchers (JCJC) grant from the French National Research Agency (ANR).
  
\section*{Appendix: Asymptotic estimates for the elastic moduli}

Here we present asymptotic estimates for the Frank elastic moduli, $K_{1}$ (splay),  $K_{2}$  (twist), $K_{3}$ (bend) and $K^{\ast}$,  that feature in the director fluctuation spectra. The corresponding microscopic expressions are very similar to \eq{k2}.  Fixing the reference director orientation along the $x$-axis of the laboratory frame (see \fig{fig0}(a)) we formulate \cite{allenevans}
 \begin{align}
\beta K_{1} &= \frac{ \rho_{0}^{2} }{2}\langle \langle w_{1z} w_{2z}  M_{2} \dot{f}_{0}( \bwa  )\dot{f}_{0}(\bwb  ) \rangle \rangle, \nonumber \\ 
\beta K_{3} &= \frac{ \rho_{0}^{2} }{2}\langle \langle  w_{1z} w_{2z}  M_{2}^{(x)} \dot{f}_{0}( \bwa  )\dot{f}_{0}(\bwb  )  \rangle \rangle, \nonumber \\
\beta K_{*} &= \frac{ \rho_{0}^{2} }{2} \langle \langle w_{1x} w_{2x}  M_{2} \dot{f}_{0}( \bwa  )\dot{f}_{0}(\bwb  )  \rangle \rangle,
\end{align}
where 
\begin{align} 
M_{2}^{(x)} & = -\int d \bdr (\bdr \cdot  \bx)^{2} \Phi(  \bdr   , \bwa , \bwb ) \nonumber \\ 
& = \frac{1}{6} L^{4}D [ (\bwa \cdot \bx)^{2} + (\bwb \cdot \bx)^{2} ].
\end{align} 
The elastic constants depend primarily on the achiral hard core of the particles and are assumed unaffected by the weak chirality imparted by the chiral potential \eq{uchir}.  For strongly elongated hard rods Onsager's theory can be invoked. Approximate analytical results can be obtained by employing a simple Gaussian test function for the ODF applicable to the regime where the local degree of nematic order is asymptotically large. The details of the analysis are outlined in Odijk's paper \cite{odijkelastic} and the asymptotic expressions for the elastic moduli are as follows
\begin{align}
\beta K_{1}D &\sim \frac{7}{32}c_{0},  \hspace{1cm} \beta K_{2}D \sim \frac{7}{96} c_{0}, \hspace{0.5cm} (K_{1} = 3 K_{2}) \nonumber \\ 
 \beta K_{3}D &\sim  \nonumber \frac{\pi}{48} c_{0}^{3}, \hspace{1cm} \beta K_{*}D \sim  \frac{5 \pi}{48} c_{0}^{3}, \hspace{0.5cm} (K_{*} = 5 K_{3})
 \end{align}
where $c_{0} = \rho_{0} L^{2} D$ denotes a dimensionless rod concentration.

\bibliographystyle{apsrev}
\bibliography{refs}

\end{document}